\documentclass[11pt,twoside]{article}

\usepackage{asp2014}

\aspSuppressVolSlug
\resetcounters

\begin{document}

\title{Citizen COmputing for Pulsar Searches: CICLOPS}

\author{
Matteo Bachetti,$^1$ 
Maura Pilia,$^1$ 
Stefano Curatti,$^2$
Giada Corrias,$^3$
Andrea Addis,$^2$
Claudia Macci\`o,$^1$
Daniele Muntoni,$^2$
Viviana Piga,$^1$
Nicol\`o Pitzalis,$^2$ and
Alessio Trois$^1$}
\affil{$^1$INAF-Osservatorio Astronomico di Cagliari, Selargius (CA), Italy; \email{matteo.bachetti@inaf.it}}
\affil{$^2$Infora Soc. Coop., Cagliari, Italy}
\affil{$^3$Universit\`a degli Studi di Cagliari, Cagliari, Italy}

\paperauthor{Matteo Bachetti}{matteo.bachetti@inaf.it}{0000-0002-4576-9337}{INAF-Osservatorio Astronomico di Cagliari}{}{Selargius}{CA}{09047}{Italy}
\paperauthor{Maura Pilia}{maura.pilia@inaf.it}{0000-0001-7397-8091}{INAF-Osservatorio Astronomico di Cagliari}{}{Selargius}{CA}{09047}{Italy}
\paperauthor{Stefano Curatti}{stefano.curatti@infora.it}{}{Infora Soc. Coop.}{}{Cagliari}{CA}{09122}{Italy}
\paperauthor{Giada Corrias}{g.corrias23@studenti.unica.it}{0000-0002-4335-1682}{Universit\`a degli Studi di Cagliari}{Dipartimento di Pedagogia, Psicologia, Filosofia}{Cagliari}{CA}{09124}{Italy}
\paperauthor{Andrea Addis}{andrea.addis@infora.it}{}{Infora Soc. Coop.}{}{Cagliari}{CA}{09122}{Italy}
\paperauthor{Claudia Macci\`o}{claudia.maccio@inaf.it}{}{INAF-Osservatorio Astronomico di Cagliari}{}{Selargius}{CA}{09047}{Italy}
\paperauthor{Daniele Muntoni}{daniele.muntoni@infora.it}{}{Infora Soc. Coop.}{}{Cagliari}{CA}{09122}{Italy}
\paperauthor{Viviana Piga}{viviana.piga@inaf.it}{}{INAF-Osservatorio Astronomico di Cagliari}{}{Selargius}{CA}{09047}{Italy}
\paperauthor{Nicol\`o Pitzalis}{nicolo.pitzalis@infora.it}{}{Infora Soc. Coop.}{}{Cagliari}{CA}{09122}{Italy}
\paperauthor{Alessio Trois}{alessio.trois@inaf.it}{}{INAF-Osservatorio Astronomico di Cagliari}{}{Selargius}{CA}{09047}{Italy}



\begin{abstract}
Most periodicity search algorithms used in pulsar astronomy today are highly efficient and take advantage of multiple CPUs or GPUs. The bottlenecks are usually represented by the operations that require an informed choice from an expert eye. 
CICLOPS is a citizen science project designed to transform the search for pulsars into an entertaining 3D video game. We build a distributed computing platform, running calculations with the user's CPUs and GPUs and using the unique human abilities in pattern recognition to find the best candidate pulsations.
\end{abstract}




\section{Introduction}

Pulsars are neutron stars (hereafter NS), whose characteristic pulsed signal is due to the emission of a beam of radiation that swipes the sky like a lighthouse as the star rotates.
Today, we  know a few thousands of them. The first was discovered in 1967 by Jocelyn Bell \citep[Nobel Prize to Hewish and Ryle in 1974]{hewish_observation_1968}.
Due to their extreme density, with a solar mass contained in a radius of $\sim10$\,km, they are considered an excellent test bed for theories describing matter at extreme densities. 
Moreover, their extreme inertia makes them very stable rotators, and their pulsed signals often reach a precision comparable to atomic clocks,
making them spectacular tools for the study of General Relativity (GR).
The first (indirect) detection of gravitational waves was performed by studying the decay of the orbit of a double NS system, using one of the two as a clock \citep[Nobel Prize to Hulse and Taylor in 1992]{taylor_new_1982}.
Today, pulsars in binary systems with other NSs or white dwarves are routinely used to precisely test the prediction of GR \citep[e.g.][]{lyne_double-pulsar_2004, archibald18}.

Most pulsars' single pulses are virtually undetectable due to background noise, but their periodicity allows a detection using power density spectra (PDS) or other periodograms.
However, there are a few sources that give off detectable single pulses, whose study is a fundamental tool to investigate the emission mechanism of pulsars.
It was during a search for these single pulses that astronomers stumbled upon a new phenomenon \citep{lorimerBrightMillisecondRadio2007}: the so-called \textit{Fast Radio Bursts} (FRBs).
They are millisecond-duration bright single pulses, of yet unknown origin \citep{plattsLivingTheoryCatalogue2019} from faraway galaxies \citep{petroff_frbcat:_2016,ravi_observed_2019}.
The energies at play are extreme, orders of magnitude higher than the most powerful known single pulses from NSs. 

CICLOPS aims at detecting pulsars and FRBs by transforming ``candidate signals'' found in automatic searches into 3D objects that the players can interact with in a game environment.
This allows us to borrow computational resources from the gamer to perform the calculations needed for the analysis, and to use the unique pattern recognition capabilities of the human eye to pick ``interesting'' or ``different'' signals that might hide a pulsar, a FRB or even something never observed before.

\section{Pulsar search fundamentals}
\begin{figure}
\centering
\includegraphics[width=0.48\textwidth]{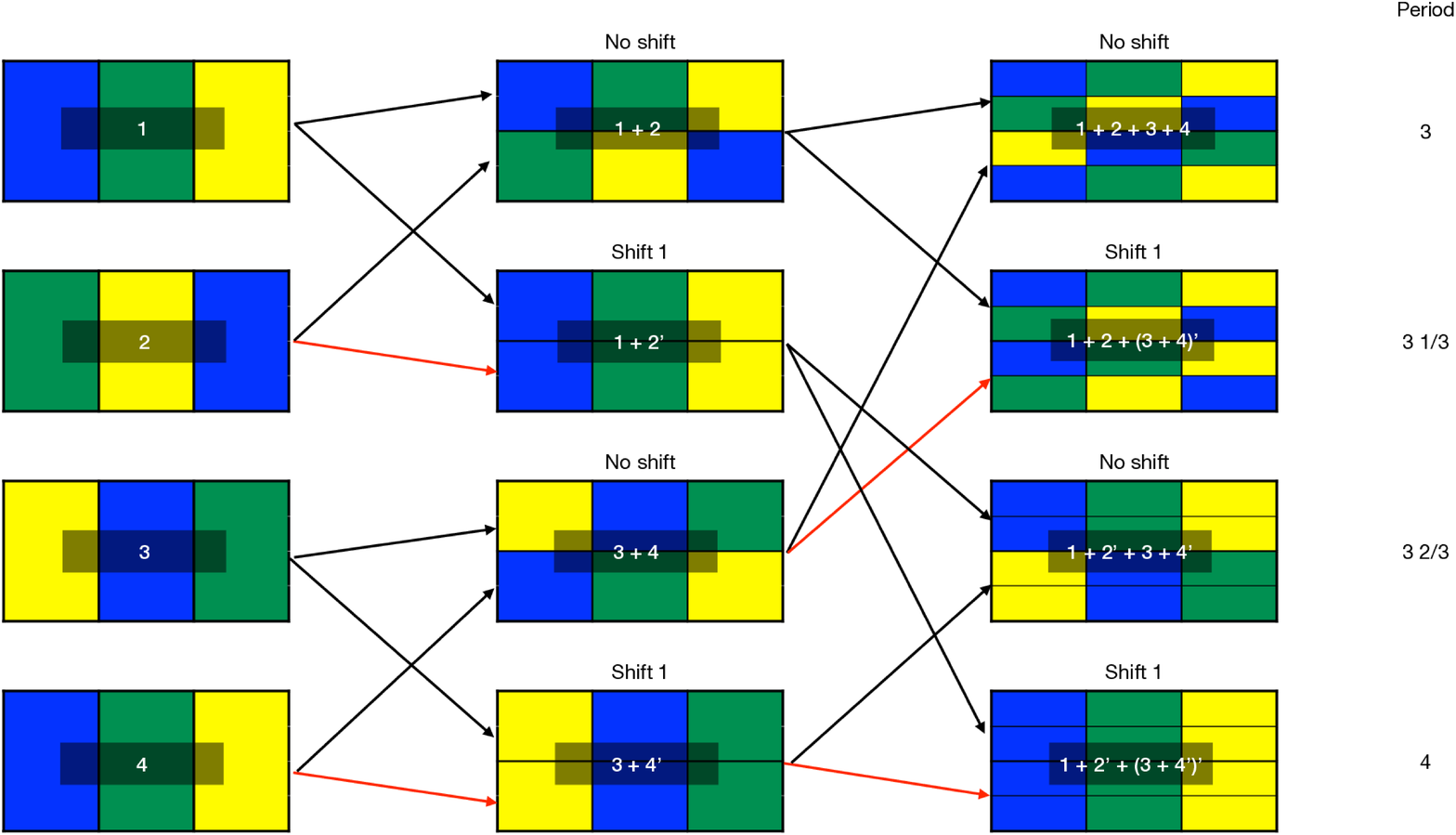}
\includegraphics[width=0.40\textwidth]{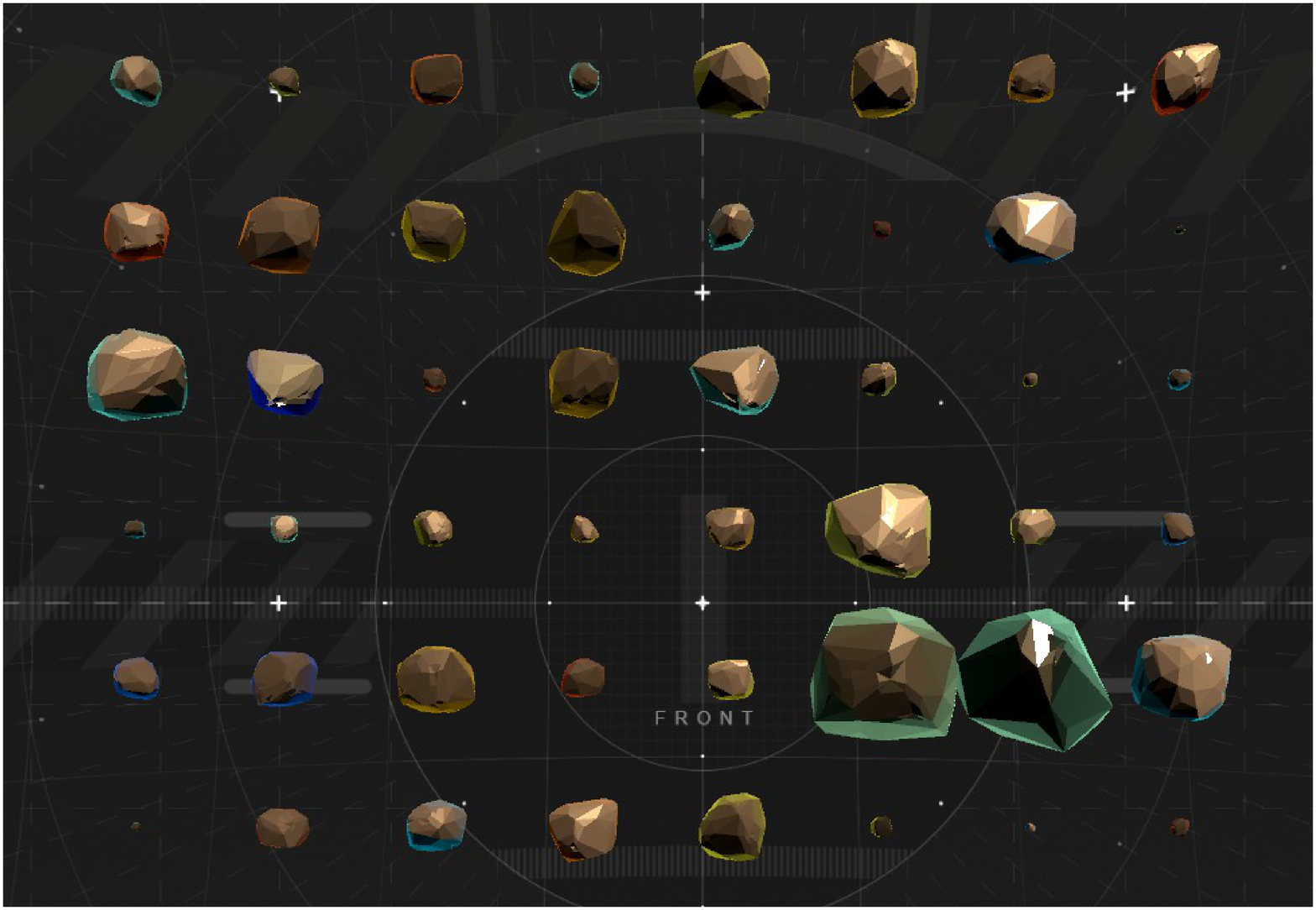}
\caption{(Left) Schematics of the Fast Folding Algorithm.
Phases of the pulsations are color-coded, so that the peak of the pulse is always in the same color (i.e. blue). 
The analysis is performed by summing blocks of data by pairs, with (black arrows) or without (red) a ``rolled shift'' of the second block during the sum.
As can be seen from the image, only one combination of such shift-and-sum aligns the pulse in the final sum. 
(Right) Candidates from an FRB search, transformed in game objects by CICLOPS. Different dimensions, colors and glow intensity/color indicate different values for statistical indicators of the pulse profile.
\label{fig:ffa}}
\end{figure}

Radio signals propagate through a ionized medium.
This column of free electrons along the line of sight between the source and the observer slows down the speed of light in the medium, in particular at low frequencies.
The \textit{dispersion} delay is calculated as $\tau_{\rm DM} \approx 4.149 {\rm DM} \nu^{-2}$\,ms, where DM (\textit{dispersion measure}) measures the amount of free electrons along the line of sight in parsecs per cm$^3$ and $\nu$ is the observing radio frequency in MHz.
In a pulsar search, the data are converted into many time series, each ``dedispersed'' for a blind trial DM: one applies a delay to the data in different radio frequency channels, correcting for the delay at the trial DM, then sums the signal in all channels to obtain a single time series.
Each dedispersed time series is usually analyzed through a Power Density Spectrum (PDS; \citealt{vanderklisFourierTechniquesXray1989} for a review). 
Since the beginnings of pulsar astronomy, a strong development of Fourier Transform algorithms \citep[e.g.][]{ransomFourierTechniquesVery2002a} has addressed the typical problems with pulsar searches. 
When we find a candidate pulsation, we usually confirm it through \textit{Epoch Folding} (EF): 
given a candidate pulsation period $p$ and a light curve, EF consists of cutting the time series in intervals of length $p$ and summing all sub-intervals.
If a signal is pulsed and the cut is done exactly, as one averages more and more sub-intervals the random measurement noise will be beaten down and the shape of the pulsation will appear.
The \textbf{Fast folding algorithm} (FFA; see Fig. 1; \citealt{staelin_fast_1969}) is an efficient way to use the EF for pulsar searches.
Whilst EF would be an O($N^2$) operation, where $N$ is the number of samples in our time series, the FFA is O($N\log N$). 
Even though the FFA is not nearly as fast as the FFT, it has recently shown its yet uncovered potential for pulsar discoveries \citep{morelloOptimalPeriodicitySearching2020}.
The FFA is the core of CICLOPS' pulsar search strategy.

\section{The CICLOPS project}
CICLOPS is a \textit{Human Computation} (HC) project, a rather new research area aimed at exploiting human intelligence in specific tasks where it is still more efficient than the most powerful and refined computing algorithms \citep{von2005human}.
Human beings are often more effective than computer algorithms at classifying, generalizing and discriminating also through sudden intuitions and/or creativity.
Specifically, CICLOPS is a \textbf{Game With A Purpose} (GWAP). In this branch of HC, scientific computational-heavy problems are wisely masked through gamified activities.
%
%
The careful gamification process \citep{articleRobsonGamification} of these analysis activities motivates players to devote time and energy to them through the design and development of game mechanics (objectives, rewards) and dynamics (how the user will interact with the mechanics).

\textbf{Platform architecture.}
The chosen technology has to fulfill three different needs: computational needs, technology availability, opportunities for human computation.
To be compatible with a game, we need to perform complex calculations in a short time, and have a stable broadband internet connection for the data exchange with remote servers. 
According to the Newzoo Consulting 2020 Global Games Market Report, mobile devices have a market share of 48\%. However, they have very low computational capabilities. The share of gaming consoles (28\%) is split between incompatible platforms (PS4, Xbox, Switch). We chose then to target PCs, because they represent a single platform with a share of 23\%, supporting in addition a wider range of interface devices (e.g. for Virtual Reality). 
As a development framework, we selected the Unity Game Engine, that has a wide support for VR development and enables strongly optimized code execution thanks to the DOTS framework.

Two main actors cooperate to complete an accurate and effective analysis in CICLOPS: volunteers/players and researchers.
In the first category we consider users that install and use the game into their devices. The current client application for CICLOPS is a VR game, but more client applications can be developed in the future. 
In the second category we include accredited scientific staff of the project. Each of them can upload classification jobs and monitor how the platform distributes them to volunteers and how volunteers are responding in terms of solutions provided.

The platform exploits an abstract client application model, with users, games and scores. This abstraction exists only at platform level and is designed as an instrument to evaluate user skills and reliability in the specific task of classification. The abstract game decouples the skills that are typically required for good entertainment from the skills that are actually needed to improve efficiency and precision in the classification process. These skills are evaluated also with a number of test datasets mixed with research-quality datasets. 
%
%
A score is assigned to jobs based on how many users flag game objects containing positive clues of the presence of a pulsar, weighted for the users' measured skill score.

\textbf{Conversion of statistics into game objects.}
The conversion of statistics into game objects is one of the most complex aspects in the development of CICLOPS. 
We face different obstacles: the data source, the game design, the medium (both the VR HMD and the game engine) all influence in some way the resulting game object. 
Before making further considerations, we need to be aware that in game development resources are scarce: any excessive use of the processing power available can potentially bring to a drop in frames per second below an acceptable threshold; this brought the need for the careful selection of a small set of statistics that could characterize a wide range of signals.
The statistical parameters have then to be represented visually, in a 3D and immersive environment, using them to drive the game object rendering.
To face these issues, after the identification of a set of variables we could extract from the candidate profile, we identified a set of game object's parameters we could influence with those values, like size, rotation speed, surface color. After that, we proceeded empirically to make the correct combination between the two using a software we developed specifically for the task (see Fig. 1). 
The scientific part of the analysis is wrapped in a more complex and structured game, with the objective of adding depth in the story, and making it entertaining and engaging, to ensure more time spent on it. During game activities not related to human computation the users' PCs computational capabilities are used to perform complex statistical calculations.


\section{Conclusions and future work}
CICLOPS is a game designed to allow the search for pulsars and FRBs. 
The game is reaching a demo stage, and future developments aim at the construction of a user base of early-adopters who, through a feedback system, will help in the development of the most critical aspects of the project: game mechanics and 
human computation. These critical aspects cannot be developed without extensive testing with a wide user base. Also, building a community could help sustain at least partially the development costs.





\acknowledgements 
This work was supported by \textit{POR FESR Sardegna 2014 - 2020 Asse 1 Azione 1.1.3} 
(code RICERCA\_1C-181), call for proposal 
``Aiuti per Progetti di Ricerca e Sviluppo 2017'' 
managed  by Sardegna Ricerche.
\bibliography{O7-45}  

\begin{thebibliography}{}
\expandafter\ifx\csname natexlab\endcsname\relax\def\natexlab#1{#1}\fi
\expandafter\ifx\csname url\endcsname\relax
  \def\url#1{\texttt{#1}}\fi
\expandafter\ifx\csname urlprefix\endcsname\relax\def\urlprefix{URL }\fi
\providecommand{\eprint}[2][]{\url{#2}}

\bibitem[{{Archibald} et~al.(2018)}]{archibald18}
{Archibald}, A.~M., et~al. 2018, \nat, 559, 73. \eprint{1807.02059}

\bibitem[{Hewish et~al.(1968)Hewish, Bell~Burnell, Pilkington, Scott, \&
  Collins}]{hewish_observation_1968}
Hewish, A., Bell~Burnell, S.~J., Pilkington, J., Scott, P., \& Collins, R.
  1968, Nat., 217, 709

\bibitem[{Lorimer et~al.(2007)Lorimer, Bailes, McLaughlin, Narkevic, \&
  Crawford}]{lorimerBrightMillisecondRadio2007}
Lorimer, D.~R., Bailes, M., McLaughlin, M.~A., Narkevic, D.~J., \& Crawford, F.
  2007, Science, 318, 777

\bibitem[{Lyne et~al.(2004)}]{lyne_double-pulsar_2004}
Lyne, A.~G., et~al. 2004, Science, 303, 1153

\bibitem[{{Morello} et~al.(2020){Morello}, {Barr}, {Stappers}, {Keane}, \&
  {Lyne}}]{morelloOptimalPeriodicitySearching2020}
{Morello}, V., {Barr}, E.~D., {Stappers}, B.~W., {Keane}, E.~F., \& {Lyne},
  A.~G. 2020, \mnras, 497, 4654. \eprint{2004.03701}

\bibitem[{Petroff et~al.(2016)}]{petroff_frbcat:_2016}
Petroff, E., et~al. 2016, PASA, 33, e045

\bibitem[{Platts et~al.(2019)Platts, Weltman, Walters, Tendulkar, Gordin, \&
  Kandhai}]{plattsLivingTheoryCatalogue2019}
Platts, E., Weltman, A., Walters, A., Tendulkar, S.~P., Gordin, J. E.~B., \&
  Kandhai, S. 2019, Physics Reports, 821, 1

\bibitem[{Ransom et~al.(2002)Ransom, Eikenberry, \&
  Middleditch}]{ransomFourierTechniquesVery2002a}
Ransom, S.~M., Eikenberry, S.~S., \& Middleditch, J. 2002, AJ, 124, 1788

\bibitem[{Ravi(2019)}]{ravi_observed_2019}
Ravi, V. 2019, Monthly Notices of the Royal Astronomical Society, 482, 1966

\bibitem[{Robson et~al.(2015)Robson, Plangger, Kietzmann, McCarthy, \&
  Pitt}]{articleRobsonGamification}
Robson, K., Plangger, K., Kietzmann, J., McCarthy, I., \& Pitt, L. 2015,
  Business Horizons

\bibitem[{Staelin(1969)}]{staelin_fast_1969}
Staelin, D.~H. 1969, in Proceedings of the {IEEE} (NRAO, Charlottesville, VA,
  USA), 724

\bibitem[{Taylor \& Weisberg(1982)}]{taylor_new_1982}
Taylor, J.~H., \& Weisberg, J.~M. 1982, ApJ, 253, 908

\bibitem[{{van der Klis}(1989)}]{vanderklisFourierTechniquesXray1989}
{van der Klis}, M. 1989, in Timing {{Neutron Stars}}: Proceedings of the {{NATO
  Advanced Study Institute}} on {{Timing Neutron Stars}} Held {{April}} 4-15,
  27

\bibitem[{Von~Ahn(2005)}]{von2005human}
Von~Ahn, L. 2005, Human computation. pittsburgh, ph.d thesis

\end{thebibliography}


\end{document}